\documentclass[showpacs,floatfix,twocolumn,prb,english,aps]{revtex4}
\usepackage[english]{babel}
\usepackage[usenames]{color}
\usepackage{amssymb}
\usepackage{ulem}
\usepackage{amsmath}
\usepackage{graphicx}
\usepackage{xspace}

\newcommand{\gsl}{\ensuremath{\gamma_{sl}}\xspace }

\makeatletter

\usepackage{babel}
\makeatother
\date{\today}
\begin{document}

\title{Solid-Liquid Interface Free Energy through Metadynamics simulations}

\author{Stefano Angioletti-Uberti$^{1}$}\email{sangiole@imperial.ac.uk}
\author{Michele Ceriotti$^{2}$} \author{Peter D. Lee$^{1}$} \author{Mike W. Finnis$^{1}$} 

\affiliation{$^1$Department of Materials and Thomas Young Centre, Imperial College London, Prince Consort
Road 20, SW72BP London, UK}
\affiliation{$^2$Computational Science, Department of Chemistry and Applied Biosciences, ETH Z{\"u}rich, USI Campus,
Via Giuseppe Buffi 13, CH-6900 Lugano, Switzerland}

\begin{abstract}
The solid-liquid interface free energy \gsl is a key parameter
controlling nucleation and growth during solidification and other
phenomena. There are intrinsic difficulties in obtaining accurate experimental
values, and the previous approaches to compute \gsl with atomistic 
simulations are computationally demanding. We
propose a new approach, which is to obtain \gsl from a free energy map of the phase transition
reconstructed by metadynamics. We apply this to the benchmark
case of a Lennard-Jones potential and the results  confirm the most reliable data obtained
 previously. We demonstrate several advantages of our new approach: it is simple to implement,  robust and free of hysteresis problems, it allows a rigorous and unbiased estimate
of the statistical uncertainty and  it returns a good estimate of of the thermodynamic limit 
with system sizes of a just a few hundred atoms.  It is therefore attractive for using with 
more realistic and specific models of interatomic forces. 
\end{abstract}

\maketitle
\section{Introduction}

Many important phenomena occurring in first order phase transformations,
 such as  nucleation and growth, are controlled by interfacial properties. In the theory
of solidification, one such property is the solid-liquid interfacial
free energy \gsl. This parameter controls the barrier
for nucleation of a solid in an undercooled liquid and the transitions between planar, cellular and dendritic
growth regimes in metals, which in turn governs their final microstructure \cite{woodruff}. 
Despite its importance for both theoretical models and practical applications, accurate data
for the value of \gsl are not known even for the case of simple elements. There
are indeed few experimental techniques aimed at measuring this quantity (for a comprehensive
review see Ref.\cite{kelton}) and their application is complicated by the very strict control on all
experimental parameters that must be achieved to obtain accurate data. One such method for example 
involves recovering \gsl indirectly from nucleation-rate measurements \cite{kelton}. In this case,
large uncertainties in the measured values arise from the possible occurrence of heterogeneous
nucleation from very low-concentration impurities.
Reliable theoretical values would therefore be very useful.

Several methods have been developed to calculate
\gsl from \emph{in-silico} experiments with molecular dynamics, where complete control of
the ``experimental'' variables is achievable. These methods are
the Capillary Fluctuation method (CFM) \cite{cfm}, different sorts of
so-called ``cleaving'' methods 
(CM) \cite{cleavage_gilmer,cleavage} and a Classical Nucleation Theory (CNT) approach \cite{cnt}.
In CFMs the fluctuation spectrum of the interface
height is related to the interfacial stiffness $\gsl\left( \theta \right)+\gsl''\left( \theta \right)$ (where 
the second derivative is taken with respect to an angle $\theta$ defining the crystallographic orientation of the surface) through
which \gsl can be recovered by calculating $\gsl+\gsl''$  for different
interface orientations and fitting the results to an expansion  of
\gsl in  kubic harmonics \cite{kubic}. In CMs, as the name suggests, bulk solid
and liquid phases are separately cleaved and the different phases are joined to form an interface. In this way, \gsl is recovered by measuring
the work done during the process. Finally, in the CNT approach, crystalline nuclei of different sizes are inserted into a supercooled
liquid and some orientational average of  \gsl is recovered by measuring the radius of the critical nucleus $R^*$ and inserting its value
in the classical nucleation theory equation relating $R^*$ and \gsl (see for example Ref.~\cite{kaschiev}, page 46). 
We refer the interested reader to the literature for details of these calculations.
Successful applications of the  aforementioned methods have been reported
for model systems such as hard spheres \cite{cleavage_gilmer,cleavage,cfm_hard} and 
Lennard-Jones potentials \cite{cfm_lj,cleavage_lj,cnt} as well as more realistic 
semiempirical and quantum-mechanical \cite{cfm,becker2,cfm_al,cfm_mg,cvm_apl} based Embedded Atom \cite{eam}
 and Stillinger-Weber \cite{stillinger} potentials.

The CFM and CNT are derived with macroscale
approximations and thus require large simulation supercells of about $10^5$
atoms to be applicable and give accurate results. 
Cleavage methods require somewhat smaller supercells ($\approx 10^3-10^4$
atoms) but are prone to the error introduced if the sequence of simulations is not completely reversible. 
A dramatic example would be the complete solidification of the liquid while joining it to the solid
due to a large relative fluctuation in the position of the interface \cite{alfe}. 
The simulation supercell must contain a relatively large area of interface in order to avoid the occurrence of these events.
Moreover, to compute \gsl accurately and efficiently, one has to use 
a cleaving potential which mimics accurately the interactions between 
the system's particles \cite{cleavage_lj}. This must be built in an \emph{ad hoc} way for every system and can 
become cumbersome when complex many-body interactions have to be taken into account
such as for example in \emph{ab-initio}-based calculations.

These shortcomings become particularly troublesome if one consider that 
interface free energies are very sensitive to the details of the empirical potential;
 for instance, different parameterizations of EAM potentials
yield values of \gsl which vary by as much as 30\% \cite{comparison_sl}.
In order to capture the complex bonding and the unusual local environments 
present at the solid-liquid interface, and to capture accurately the anisotropy of crystalline surface energies, one must consider more sophisticated models,
which reproduce more closely the first-principles total energy.

In the present paper, we discuss a novel technique to compute \gsl which aims at being robust, efficient
and transferable, and which is a promising candidate to extend the scope of interfacial energy calculations to more complex potentials than previously treated.
Briefly, our method reconstructs a coarse-grained free energy surface (FES) using
metadynamics\cite{metadyn,parrinello-eppur}. Such a FES maps out the transition from a single phase to
the space of configurations where two phases coexist. The minimum difference in Gibbs free energy
between these two regions at the solid-liquid equilibrium temperature
is an excess free energy $G_{xs}$, which is equivalent to the interface free energy \gsl
multiplied by the area $A$ of the interface.\\

The remainder of this paper is organized as follows. In Section II
we present the thermodynamic basis and the details of the method.
In Section III we describe the computational details of our simulations.
In Section IV we show our results for a simple Lennard-Jones system
and critically discuss them in comparison with other available methods.
 We also speculate on the possibility of 
implementing this approach together with {\em ab-initio} molecular
dynamics. Finally, we summarize our main results.

\section{Methodological details}

\subsection{Thermodynamic basis}

We consider a homogeneous solid or liquid system of $N$ atoms, located in a periodically repeated supercell within an infinite system, at a
pressure $P$ and temperature $T$. Its Gibbs free energy $G$ can be written as
\begin{equation}
G_{s(l)}(P,T)=\mu_{s(l)}(P,T) N
\label{eq:gibbs-bulk}
\end{equation}
where $\mu_{s(l)}$ is the chemical potential of atoms in the solid
(liquid) phase. At the melting temperature $T_m$, the chemical potentials
in the two phases are equal
\begin{align*}
\mu_s(P,T_m)=\mu_l(P,T_m)\equiv \mu(P,T_m)\,.
\end{align*}

There exists a second state of the same system at the melting temperature,
in which solid and liquid phases coexist, separated by  macroscopically planar interfaces that 
are naturally fluctuating on the atomic scale. 
Since the chemical potential in the solid and liquid bulk phases at $T_m$ is identical,
one can write the overall Gibbs free energy as
\begin{equation}
 G_{s|l}(P,T_m)=\mu(P,T_m)N +G_{xs},
\label{eq:gibbs-interface}
\end{equation}
where an excess energy term associated with the interface has been introduced.
Such a term will be extensive with respect to the area of the 
interface, and we can write it as the product of the surface area $A$ and 
an interface free energy \gsl, i.e. $G_{xs}=A\gsl$. 

The most direct approach to the computation of \gsl is clearly
to calculate the free-energy difference between the bulk phases and 
the configurations in which planar interfaces are present, as described by Eqs.~\ref{eq:gibbs-bulk} 
and~\ref{eq:gibbs-interface} respectively.
We will obtain this free-energy difference by means of metadynamics simulations,
as  described in the next section.

\subsection{Free energy differences from metadynamics}

The use of metadynamics for reconstructing free-energy landscapes has
been the subject of many papers and we refer the reader to
the excellent review by Laio and Gervasio and references therein \cite{metadyn_review}, while we only briefly
sketch the main ideas here. Metadynamics is a simulation technique
based on non-equilibrium molecular dynamics, which is designed to reconstruct a coarse-grained free energy surface (FES) in the space of one or more collective variables $\{s_i\}$ that describe the state
of the system. Metadynamics reconstructs the FES by adding a bias potential in the
form of a Gaussian centered at a specific point in the Collective Variable (CV) 
space each time that point is visited. The mathematical
form of the bias potentials is given by 

\begin{equation}
V(\mathbf{s}_0,t)=\int_0^t w e^{-\frac{(\mathbf{s}(t')-\mathbf{s}_0)^2}{2\sigma^2}}\mathrm{d}t'
\label{eq:v-meta-continuous}
\end{equation}
which in the discrete version needed to implement the algorithm for computations
becomes
\begin{equation}
V(\mathbf{s}_0,t)=\sum_{i=0}^N w\tau e^{-\frac{(\mathbf{s}(i\tau)-\mathbf{s}_0)^2}{2\sigma^2}}.
\label{eq:v-meta-discreet}
\end{equation}
Here $\tau$ is the inverse of the frequency of deposition of the Gaussians, and
$N=t/\tau$ is the number of Gaussians accumulated up to time $t$ in the simulation.
$w$ is the deposition rate of the Gaussian functions and $\sigma$ their width. 

\begin{figure}[bpt]
\includegraphics{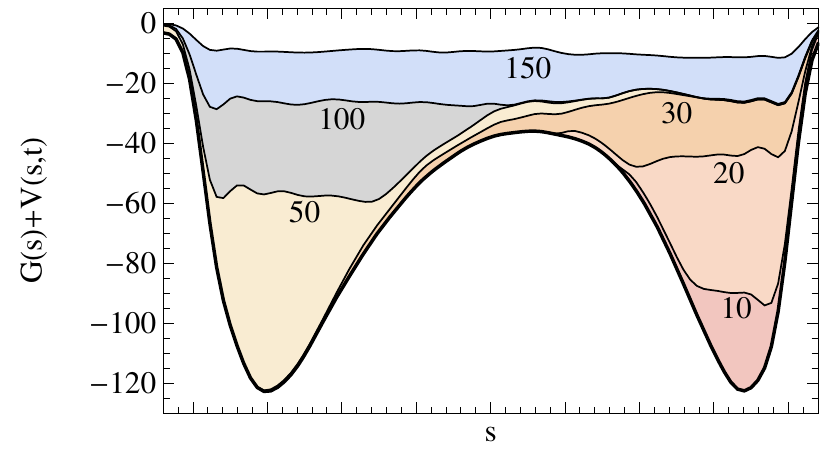}
\caption{(color online) Schematic representation of the flattening of the effective FES by the 
repulsive bias of Eq.~(\ref{eq:v-meta-discreet}), during a metadynamics simulation in 
a one-dimensional collective variable space. We show the underlying FES 
$G(s)$ and the bias accumulated at chosen times (arbitrary units).
For a sufficiently long simulation, $G(s)+V(s,t)\rightarrow constant$, so
that one can obtain an accurate estimate of the free-energy surface simply by taking
the negative of the bias.
\label{fig:metadyn}
}
\end{figure}

The bias discourages the trajectory from remaining indefinitely in the same
region of the CV space, effectively pushing the system towards the lowest-lying
free-energy barrier. Once all the relevant free energy minima have been levelled 
by the bias (see Figure~\ref{fig:metadyn}), the system becomes completely diffusive and 
wanders freely through all the possible states. 

At this stage of the simulation the accumulated bias is equal to the negative of the free energy
of the real system plus an additive constant (for a detailed analysis see Ref.\cite{metadyn_error1}).
However, there are two limitations in this original form
of metadynamics. First of all it is not clear when a metadynamics simulation should be stopped,
i.e.  when the bias has effectively compensated the underlying
free energy. Moreover, even at this point, the effective FES will have a residual
roughness of the order of the height of each individual Gaussian 
($w\tau$ in equation \ref{eq:v-meta-discreet}). 
In order to resolve these issues,
the so-called ``well-tempered'' metadynamics method \cite{welltempered_recover} has been
proposed recently by Barducci \textit{et al.}, and we use this specific type of metadynamics in our simulations.
The idea behind well-tempered metadynamics is to gradually reduce the height
of the deposited Gaussians, at a rate determined by the magnitude of the bias already present.
The expression analogous to~(\ref{eq:v-meta-continuous}) reads
\begin{equation}
\label{eq:welltempered1}
V(\mathbf{s}_0,t)=\int_0^t w e^{-\frac{V(\mathbf{s}(t'),t')}{k\Delta T}}e^{-\frac{(\mathbf{s}(t')-\mathbf{s}_0)^2}{2\sigma^2}}\mathrm{d}t'.
\end{equation}
The parameter $\Delta T$ controls how quickly the deposition rate is reduced.  Once the simulation
approaches convergence, the collective variables space will be sampled with a probability
distribution corresponding to an artificial temperature $T+\Delta T$\cite{welltempered_orig}.
Hence, the final bias accumulated during a single simulation converges to
\begin{equation}
\label{eq:welltempered2}
V(\mathbf{s}_0,t)\rightarrow-\frac{\Delta T}{\Delta T+T} G(\mathbf{s}_0)
\end{equation}
The true free energy can be recovered inverting equation~(\ref{eq:welltempered2}).

As in any free-energy calculation based on the mapping of the configurations of 
the system to a coarse-grained collective-variable space, the definition 
of CVs that can effectively distinguish between relevant states, and describe 
reliably the natural transformation path is the first, and most important
step. The primary requirement is to distinguish the solid phase from the 
liquid. With this aim, we define for every atom an order parameter $\phi$,
which depends on the relative position of its neighbors. 
The definition of $\phi$ 
\begin{equation}
 \phi(\mathbf{x}_i)=\frac{
\sum_{j\ne i} c_r\left(\left|\mathbf{x}_j-\mathbf{x}_i\right|\right) 
              c_\alpha\left(\mathbf{x}_j-\mathbf{x}_i\right)
}{
\sum_{j\ne i} c_r\left(\left|\mathbf{x}_j-\mathbf{x}_i\right|\right) 
}\label{eq:phi}.
\end{equation}
contains an angular term $c_\alpha$ to distinguish the different environments,
and some radial cutoff functions $c_r$ which are useful to guarantee 
that $\phi$ is a continuous function of all its arguments.
Note that the weighted sum of $c_\alpha$ is normalized over the 
total coordination, so that $\phi$ is relatively insensitive to fluctuations
of the density.

\begin{figure}
\includegraphics[width=0.9\columnwidth]{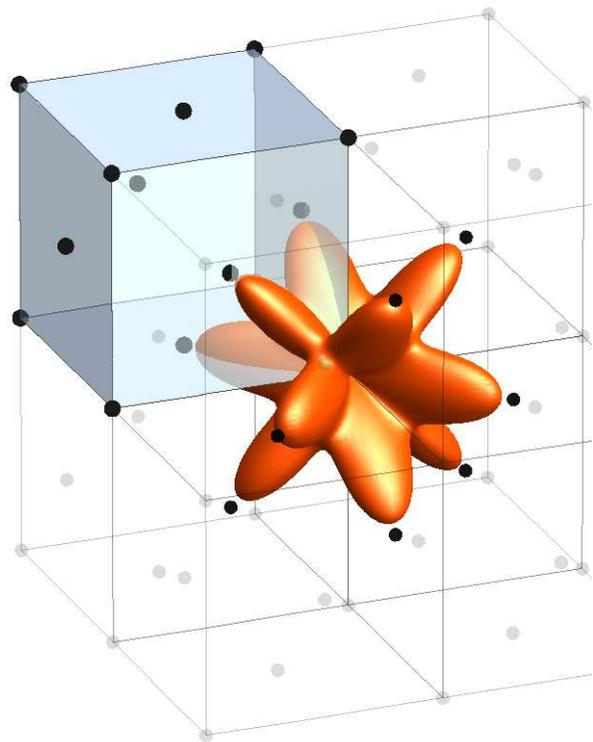}
\caption{(color online) Angular function $c_\alpha(\hat{\mathbf{x}})$ as defined in equation \ref{eq:c-alpha}.
 The function is shown as a polar plot, centered on an \emph{fcc} lattice.
$c_\alpha$ has well-defined peaks in the directions of the nearest neighbors.}
\label{fig:c-alpha}
\end{figure}

We define the angular function $c_\alpha$ as a combination of polynomials 
in Cartesian coordinates, symmetry adapted to the cubic point group: 
\begin{equation}
\begin{split}
c_\alpha(\mathbf{x})=&
 \left[x^4y^4\left(1-z^4/\left|\mathbf{x}\right|^4\right)
+y^4z^4\left(1-x^4/\left|\mathbf{x}\right|^4\right)\right.\\
&\left. +z^4x^4\left(1-y^4/\left|\mathbf{x}\right|^4\right)\right]\frac{1}{\left|\mathbf{x}\right|^8}
\end{split} \label{eq:c-alpha}
\end{equation}
We have chosen Eq.~(\ref{eq:c-alpha}) rather than more traditional 
parameters such as the so-called $Q_6$ (see e.g. \cite{q61,q62,q63}), for a number
of reasons: $c_\alpha$ has well-defined peaks for an \emph{fcc} environment
(see Figure~\ref{fig:c-alpha}), it is not rotationally invariant (and will therefore
enforce an orientation of the crystal consistent with the periodic boundaries) 
and it is relatively cheap to compute. It would possible to construct a different form of $c_{\alpha}$ if one wanted
to deal with a different crystal structure, and one simply has to rotate the function in order to specify a different crystallographic orientation of the surface. The application of a specialised, orientation-dependent order  parameter is a
key ingredient of our approach.

\begin{figure}[bpt]
\includegraphics[width=0.9\columnwidth]{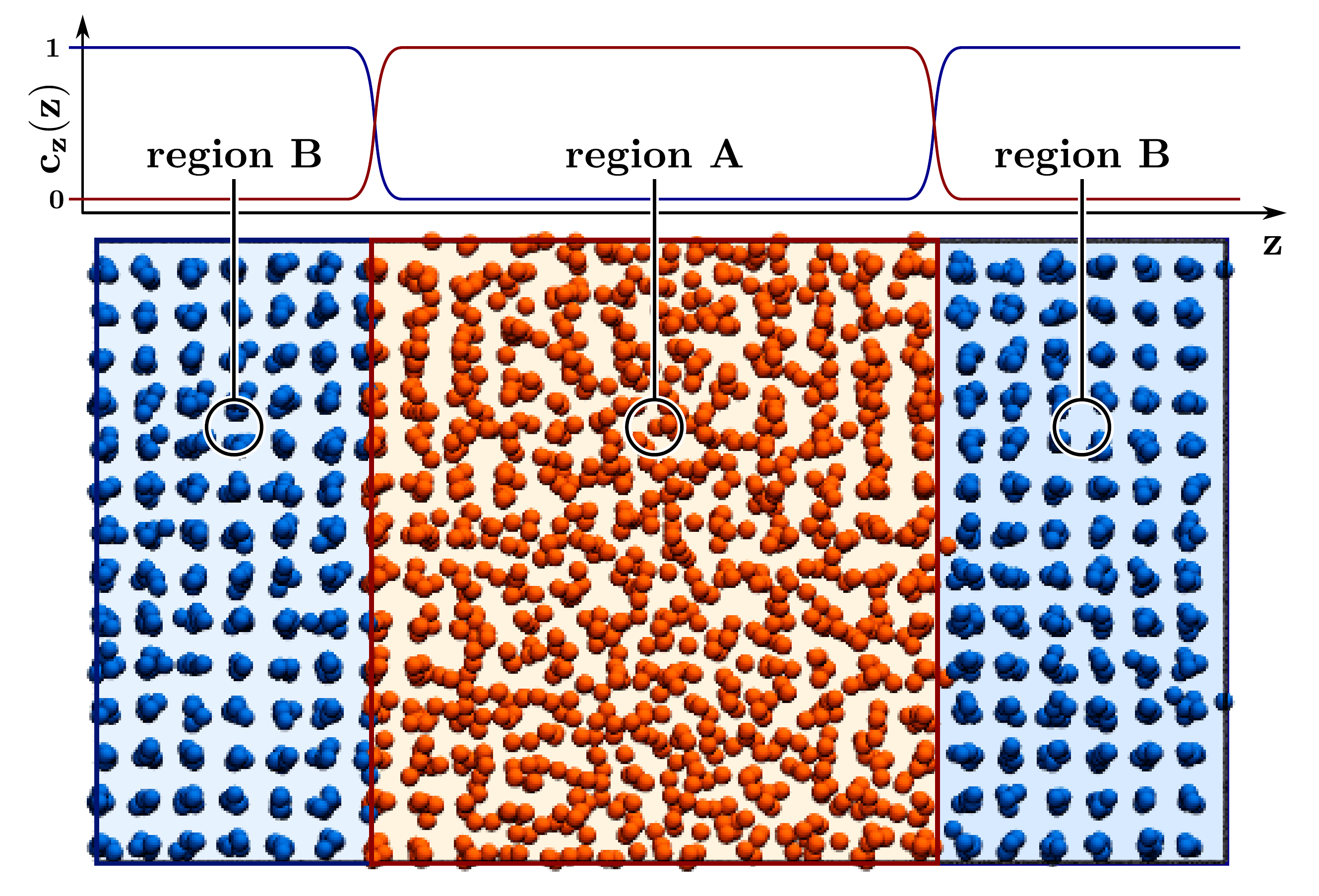}
\caption{(color online) Cutoff function used to define the regions A and B for the calculation of the
two collective variables. The function varies smoothly from 0 to 1 so as to avoid
discontinuities when atoms transit between the two regions.}
\label{fig:z-weight}
\end{figure}

The radial cutoff is defined as
\begin{equation}
\label{eq:c-r}
 c_r(r)=\begin{cases}
1 &\quad  r\le r_1 \\
0 &\quad  r\ge r_0 \\
\left[\left(y-1\right)^2\left(1+2y\right)\right] & r_1<r<r_0\\
\end{cases}
\end{equation}
where $y=(r-r_1)/(r_0-r_1)$.
The polynomial part in Eq.~(\ref{eq:c-r}) is simply a third order polynomial satisfying the 
constraints of continuity of $c_r(r)$ and its first derivative at $r_1$ and $r_0$.\\
In order to study the formation of a solid-liquid interface, one must 
then distinguish configurations where the supercell is completely solid, 
completely liquid, or partially solid and partially liquid: in the latter case, at least
two parallel interfaces will be present. 
For this purpose we divide the supercell, centered at the origin, into two regions:  we assign to region A those 
atoms having $|z|<\bar{z}$, and to region B all the others (see Figure~\ref{fig:z-weight}). 
Note that we take $\bar{z}$ to be about one fourth of the supercell length along $z$,
and we keep it constant irrespective of the fluctuations of the supercell's
size. This choice is not troublesome as long as the averages are properly normalized,
so that the value of the CVs is independent of the number of atoms 
contained in each of the regions.

Again, in order to obtain smoothly-varying CVs, we introduce a
weight function. We use the same functional form introduced for the radial cutoff;
namely, $c_z(\mathbf{x})=c_r(|z|)$, setting $r_1=\bar{z}$, $r_0=\bar{z}+\Delta z$ in Eq.~(\ref{eq:c-r}).

We finally define our CVs $s_A$ and $s_B$ by averaging the order parameters of the 
atoms located within region $A$ and $B$, respectively:
\begin{equation}
 \begin{split}
  s_A=&\sum_i \bar{\phi}(\mathbf{x}_i) c_z(\mathbf{x}_i)/\sum_i c_z(\mathbf{x}_i) \\
  s_B=&\sum_i \bar{\phi}(\mathbf{x}_i) \left[1-c_z(\mathbf{x}_i)\right]/\sum_i \left[1-c_z(\mathbf{x}_i)\right]
 \end{split} \label{eq:sasb}
\end{equation}
where
\begin{equation}
 \bar{\phi}=\frac{2288}{79} \phi -\frac{64}{79}.
\label{eq:scaled-phi}
\end{equation}
This scaling has been chosen so that $\bar{\phi}=0$ in a homogeneous
liquid and  $\bar{\phi}=1$ in a perfect \emph{fcc} solid.

\section{Computational details\label{sec:details}}

In order to evaluate our method, we decided to perform the metadynamics calculations 
with a truncated Lennard-Jones potential,
in the form used by Broughton and Gilmer \cite{gilmer}. 
Such a simple potential is inexpensive and
thoroughly studied, yet it can capture many of the relevant physical phenomena occurring in real
systems. Its solid-liquid transition, an important ingredient of our approach, has been recently 
studied by Mastny and de Pablo \cite{pablo} through a modified Wang-Landau sampling technique \cite{landau}.
Therefore, this system is ideal for a systematic study with our 
method, comparing it with other techniques and benchmarking its performance
as a function of the parameters of the simulation.

We will use Lennard-Jones units  throughout the paper. 
The zero pressure coexistence temperature for this system has been recently recalculated
and we take it to be $T_m=0.6185$ \cite{cleavage_lj,becker}. Details of the phase diagram can be found in ref.\cite{becker}.
We performed our simulations with a range of supercell sizes from $4\times 4\times 6$ 
\emph{fcc} unit cells (384 atoms) to $9\times 9 \times 20$ (6480 atoms).
The supercells were oriented with \emph{fcc} $[001]$ cell vectors parallel to the axes, with the longest  side parallel
to $z$, and were rescaled to a volume consistent with the equilibrium density of the solid at the coexistence temperature.
Atomic positions were then equilibrated at $T_m$ by performing a short molecular dynamics simulation in the 
canonical (NVT) ensemble.
This procedure was adopted in order to generate the starting configurations for the subsequent metadynamics 
simulations, which we perform instead in the isothermal-isobaric (NPT) ensemble. 
The timestep for the integration of the equations of motion, performed 
with a velocity Verlet algorithm \cite{frenkel}, was $0.004$.
This choice gave negligible drift of the conserved quantity in all our simulations.

In order to perform constant pressure simulations, variable-cell dynamics is 
implemented using a Langevin-piston barostat\cite{piston} 
and a friction of $\gamma_B=2$~ps$^{-1}$.

The presence of an interface calls for particular attention when performing 
constant-pressure simulations. In particular, one has to deal with the change in density that 
occurs when a portion of the supercell melts,  at the same time considering fluctuations in the $xy$-plane. If the $xy$-plane parameters of the supercell
are fixed, the fluctuations in the solid will be frustrated; on the other hand, if those
parameters are left free to vary, one will witness a spurious shrinking of the dimensions in the  $xy$-plane
due to surface tension whenever an interface is present. In both cases one can in principle observe
 a strain-related free energy contribution. However, this problem 
would disappear in the thermodynamic limit, hence one can just make the choice
that is more computationally convenient, and consider the resulting error 
as another finite-size effect, which can be controlled by comparing the results 
with different supercell sizes. 
We decided to let only the $z$ component free to fluctuate. In this way,
the change of volume occurring as the fraction of liquid and solid phases changes 
can be accommodated, and  even in case of complete
melting the $xy$ dimensions remain commensurate with a strain-free solid,
which makes it simpler for the system to freeze again into a defect-free solid.


Temperature control is extremely important in metadynamics
simulations, since the increase of the biasing potential 
creates a continuous supply of energy to the system, which
must nevertheless be held close to equilibrium in order to
sample the free energy correctly.
A strong local thermostat is needed, but at the same time
one must avoid overdamping, which drastically reduces the diffusion
coefficient and hence the sampling of slow, collective modes.
We therefore use a colored-noise thermostat~\cite{ceriotti,ceriotti2,gle4md}
fitted to provide efficient sampling over a broad range of frequencies, corresponding to vibrational periods
between $0.1$ and $10^3$ Lennard-Jones time units.

The metadynamics we used for the different simulations are reported in Table~\ref{tab:meta_parameters}. This Table also includes data for a number of tests using a single CV, which we describe later. We performed tests with other choices of these parameters spanning about an order of magnitude and no statistically significant changes 
were observed in the calculated value of \gsl.  The values reported, however, resulted in the best statistical
uncertainty in the final free-energy estimate.The simulations were performed
using the DL\_POLY code (version 2.18, \cite{dlpoly}),
patched to perform metadynamics using the PLUMED\cite{plumed} 
cross-platform plugin, which greatly reduces the implementation burden 
by providing a convenient framework for introducing new collective variables.\\

\begin{figure}[bhtp]
\includegraphics[width=1.0\columnwidth]{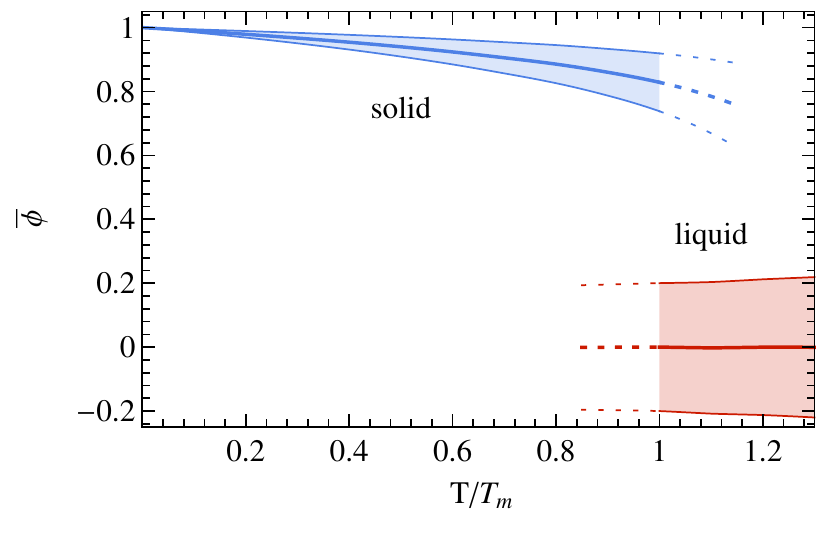} 
\caption{
(color online) TIme-averaged value of the order parameter $\bar{\phi}$ for an individual atom in the bulk solid
and liquid phases, as evaluated for the LJ system at different temperatures and across the solid-liquid 
transition.
The bounding lines delimit the range one standard deviation above and below the mean value. 
Dashed lines correspond to the values of the order parameter for metastable
solid and liquid. Note that the parameters  $r_1$ and $r_0$ for the radial cutoff~(\ref{eq:c-r})
are scaled from the values used at $T_m$ according to the changes of the equilibrium density.
\label{fig:cv_vs_temp}}
\end{figure}

\begin{table}

\begin{tabular}{ c c c c c }
\hline
&
\# atoms (cell)&
$\tau$&
$1+\frac{\Delta T}{T}$&
$w\tau$
\tabularnewline
\hline
\hline

S1 (2D) &
2352 ($7\times 7\times 12$)&
4&
60&
0.115\tabularnewline

S2 (1D) &
384 ($4\times 4\times 6$)&
4&
40&
0.037\tabularnewline

S3 (1D) &
512 ($4\times 4\times 8$)&
4&
40&
0.037\tabularnewline

S4 (1D) &
768 ($4\times 4\times 12$)&
4&
40&
0.037\tabularnewline

S5 (1D) &
1024 ($4\times 4\times 16$)&
4&
40&
0.037\tabularnewline

S6 (1D) &
1280 ($4\times 4\times 20$)&
4&
40&
0.037\tabularnewline

S7 (1D) &
1200 ($5\times 5\times 12$)&
4&
60&
0.058\tabularnewline

S8 (1D) &
2352 ($7\times 7\times 12$)&
4&
120&
0.115\tabularnewline

S9 (1D) &
6480 ($9\times 9\times 20$)&
4&
205&
0.191\tabularnewline

\hline
\end{tabular}
\caption{
Metadynamics parameters for different simulations (1 and 2 dimensional) and different supercell sizes. 
 $\Delta T$ has been chosen such that $k(\Delta T +T)\approx \gsl A$ for 
every size. An order-of-magnitude estimate of $\gsl$
suffices for this purpose. $\tau$ was chosen
so as to observe the first solid-liquid transition at about half of the total simulation time, 
and $\sigma$ was set to $0.03$, which is of the order of the thermal fluctuations of 
the CVs in an unbiased simulation.
\label{tab:meta_parameters} }
\end{table}
When performing simulations at $T > 0K$, thermal fluctuations induce some disorder in the solid
and the scaled order parameter $\bar{\phi}$ deviates from the value predicted for the ideal \emph{fcc} crystal.
In figure \ref{fig:cv_vs_temp} we report the time-averaged order parameter $\left<\bar{\phi}\right>$ and its
fluctuations for a single atom in the bulk phases. 
At the coexistence temperature, the average for the 
liquid oscillates around zero, while for the solid $\left<\bar{\phi}\right>\approx 0.83$.
We note that even for an individual atom the order parameter can distinguish very clearly between
the two phases at the melting temperature.

The parameters entering the radial cutoff function $c_r$ have been chosen to be 
$r_0=1.5$ and $r_1=1.2$, so as to encompass the typical first-neighbour distances in both 
Lennard-Jones solid and liquid at $T=T_m$. 
In order to prevent $s_A$ and $s_B$ from visiting irrelevant configurations, corresponding
to an order parameter far from its mean value in either liquid or solid, we have applied
a lower and upper wall on both CVs \cite{metadyn_review} 
in the form 
\begin{equation}
\label{eqn:vwall}
V_{wall}(s)=k\left(\frac{s-s_{limit}}{\epsilon}\right)^n
\end{equation}
with 
$k=50$, $\epsilon=0.01$, $n=4$ and $s_{limit}=-0.15$ and $0.95$ for the lower and upper wall respectively, which introduces
 a restraining potential for $s_A,s_B<-0.15$ and $s_A,s_B>0.9$. At the same time, $V_{wall}$ is set to be zero inside this interval
 and thus cannot interfere with the dynamics of the system in this region of CV space.

\begin{figure*}[bhtp]
\includegraphics{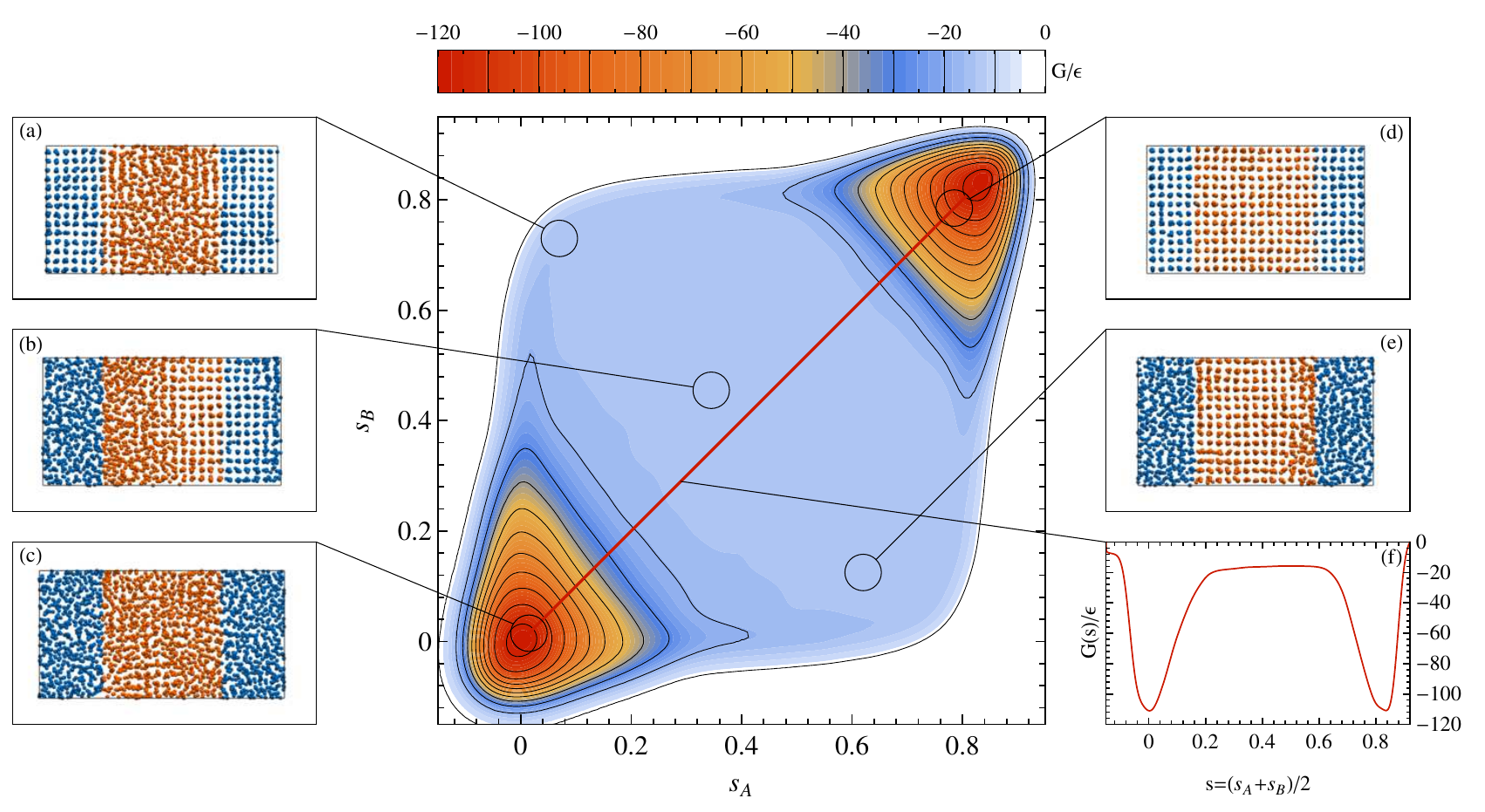}
\caption{(color online) 2D FES reconstructed by well-tempered metadynamics, together with selected 
snapshots of configurations obtained during the simulation. Atoms participating in $s_A$ are colored in orange, 
those in $s_B$ in blue, and the region of CV space corresponding to each snapshot is marked.
The negative peaks in the FES clearly correspond to the two single-phase regions. They are separated by a very wide plateau, corresponding to the presence of  well-defined interfaces between
 solid and liquid phases at various different positions relative to the $A/B$ partition (insets (a), (b), (e)). 
In inset~(f)  we report the projection of the FES along the single CV $s=(s_A+s_B)/2$.
\label{fig:fes2d}}
\end{figure*}

\section{Results and Discussion}\label{sec:results}

\subsection{Qualitative analysis of the FES}

Ideally the FES should be symmetric about the line $s_A = s_B$. Moreover,
as calculations are performed at $T_m$, one should observe the occurrence of two minima with the same free energy,
at the values of the CVs corresponding to the single bulk phases (either solid or liquid).
The expected behavior is clearly exhibited by the calculated FES, reported in
Figure~\ref{fig:fes2d} for a 7x7x12 supercell, where we show the free-energy landscape 
together with some snapshots corresponding to different values of the CVs.
The combination of CVs $s=(s_A+s_B)/2$ corresponds roughly to the average of $\bar{\phi}$ over
the whole box, and distinguishes between configurations with different proportions of solid and liquid phases.
It can be used as a convenient reaction coordinate (see Figure~\ref{fig:fes2d}(f) for the FES projected on $s$).
As expected, two wells occur with minima at the complete solid and liquid states, separated by a rather flat region, 
whose height above the two minima corresponds to the interfacial free energy.

The fact that the two wells should have the same depth can be used 
to check that the simulation temperature is indeed the melting temperature.
This is a significant advantage of our method, since knowledge of the melting temperature is a
prerequisite of all the different methods described in the literature to calculate \gsl. Both the CFM and 
CM, being based on coexistence simulations, could suffer from major irreversible changes
leading to complete solidification/melting of the simulation cell if not performed at the correct temperature, and the data gathered during the simulation would not serve its purpose. Also the CNT method needs the correct value for the melting temperature in order
to calculate the nucleation barrier from which \gsl is recovered. 
Our method, by contrast, is still effective even if the simulation temperature is slightly off the actual $T_m$. 
Clearly, an error will be introduced, since \gsl is estimated from equation \ref{eq:gibbs-interface} 
which is satisfied exactly only at $T=T_m$. However, our method is very robust, in the sense that it tells us both whether
such an error occurs and give us the sign and an estimate of the magnitude of the correction.
We will discuss this issue further when addressing finite-size effects.

The two-dimensional FES is rather constant along the line of points equidistant between the two wells,  in the direction of the orthogonal combination of CVs $\bar{s}=(s_A-s_B)/2$,
since this variable describes the position of the two phases with respect to the partitioning of the cell along $z$
(see snapshots a, b and e in Figure \ref{fig:fes2d}).
As we will comment further below, our CVs are effective because there exist no metastable phases of the LJ potential
correspond to values of the order parameter $\bar{\phi}$ 
between $\bar{\phi}_{l}$ and $\bar{\phi}_s$, and this is likely to be the case for any potentials describing a \emph{fcc} crystal. Hence, when moving away from the perfect liquid(solid) bulk value,
any homogeneous variation of the order parameter would be too energetically costly, and 
the system instead induces some order(disorder) in the form of small clusters, 
slowly increasing the free energy (region in orange in Figure~\ref{fig:fes2d}).
Because of the elongated aspect ratio of the supercell, as soon as enough liquid(solid) phase is present,
the most favourable configuration corresponds to the presence of two interfaces perpendicular to the $z$
axis. The fact that we observe a solid$\leftrightarrow$liquid transition via the growth of an individual 
cluster suggests that a very similar approach can be used to study the nucleation process itself. This
idea will be explored in future work.

As the time needed by metadynamics to reconstruct the FES is an exponential function
of the dimensionality of the coarse grained space, one might wonder if it is possible to 
speed up calculations by using a single CV. We explored this possibility as follows. Rather than using $s=(s_A+s_B)/2$, we kept 
a two-dimensional description, but performed metadynamics on $s_A$ alone, while 
 atoms in region B  are constrained in order to maintain this region of the supercell in 
a solid state (i.e we apply a restraining lower wall potential which is a function of $s_B$, and introduces a 
penalty in the enthalpy whenever $s_B$ deviates too much from $\bar{\phi}_s$. 
The values of the parameters entering in the restraining potential, whose form is described in Section
 \ref{sec:details}, are $k=50$, $\epsilon=0.01$, $n=4$ and $s_{limit}=0.7$.
This forces region B to remain solid, while region A can sample both solid and liquid phases.
In this case, the FES should show a minimum for $s_A\approx \bar{\phi}_s$ where the supercell is completely in a solid phase
and a maximum where $s_A\approx \bar{\phi}_l$  i.e. when two interfaces are present.
Again the difference between these two values is the interface excess energy. Figure~\ref{fig:meta-conv} (inset (a))
shows the 1D FES reconstructed in this way at different simulation times.
The use of a single CV does not have any adverse influence on the calculated
value of \gsl, as we will show in Section~\ref{sec:error}, and the use of this simpler form of metadynamics is fully justified
for our purposes.

\begin{figure}[bpt]
\includegraphics{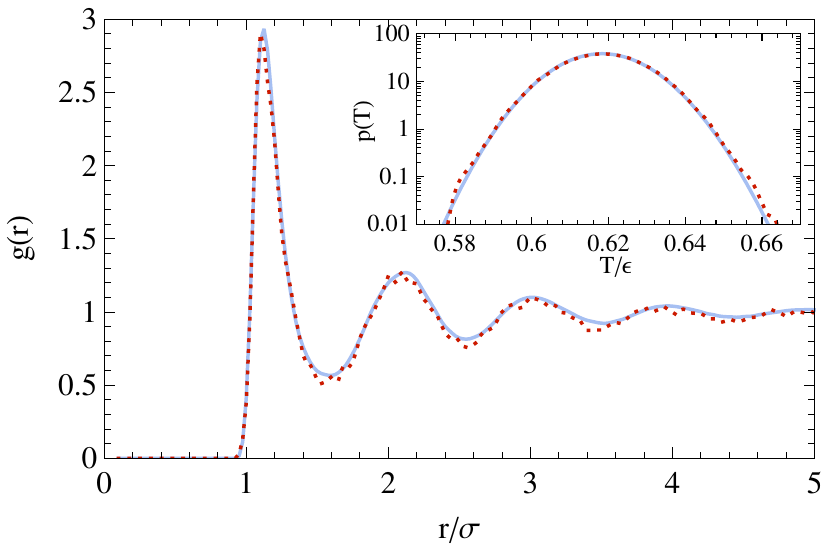}
\caption{(color online) Radial pair correlation function $g(r)$ for the liquid in the presence of an interface during
our metadynamics simulations (red, dashed) and for a normal molecular dynamics simulation of a bulk liquid (blue). 
It is clear that the two curves are very similar, thus ruling out quantitatively the presence of metastable
structures during our simulation. In the inset, the kinetic energy distribution during
our simulation is plotted in comparison to that expected for a canonical ensemble at $T=T_m$. 
Again the two are very close demonstrating that the metadynamic bias does not induce any systematic deviation
from the correct ensemble, and that quasi-equilibrium conditions hold.}
\label{fig:eq-test}
\end{figure}
 
A necessary condition for metadynamics to reconstruct the coarse-grained
free energy of the system in a meaningful way is that all the important states and the barriers
between them are effectively reached many times during the simulation. 
Moreover, one has to make sure that quasi-equilibrium conditions hold,
which can be monitored by checking temperature and structural relaxation
of the system.

In order to check that the system effectively performs many transitions
between the single-phase and the two-phase states, 
we verified that the CVs oscillate several time between their value in the liquid
and solid phases. Moreover, we also visually check that the system actually
performs these transitions by printing snapshots of the atomic positions and visualising
them using the VMD software\cite{vmd}.
Quasi-equilibrium conditions hold very accurately, as demonstrated
from the inset (b) of Figure~\ref{fig:eq-test} where we show the velocity distribution function
 compared to its analytical equilibrium value. 
The radial pair correlation function $g(r)$ of the liquid portion of interfacial 
configurations (Figure~\ref{fig:eq-test}) agrees well with the one
computed for the bulk liquid in an unbiased run, which is a further
confirmation that our simulation strategy does not introduce spurious
structural effects. The  $g(r)$ distribution is a sensitive measure of the short-range 
order present in the liquid, and any extra structuring would have been clearly 
detected as a shift of the peak positions or shapes, which does not happen here.
The absence of artefacts has also been checked by visual inspection of snapshots
of the atomic configurations along the metadynamics trajectory.

\begin{figure}[bpt]
\includegraphics[width=1.0\columnwidth]{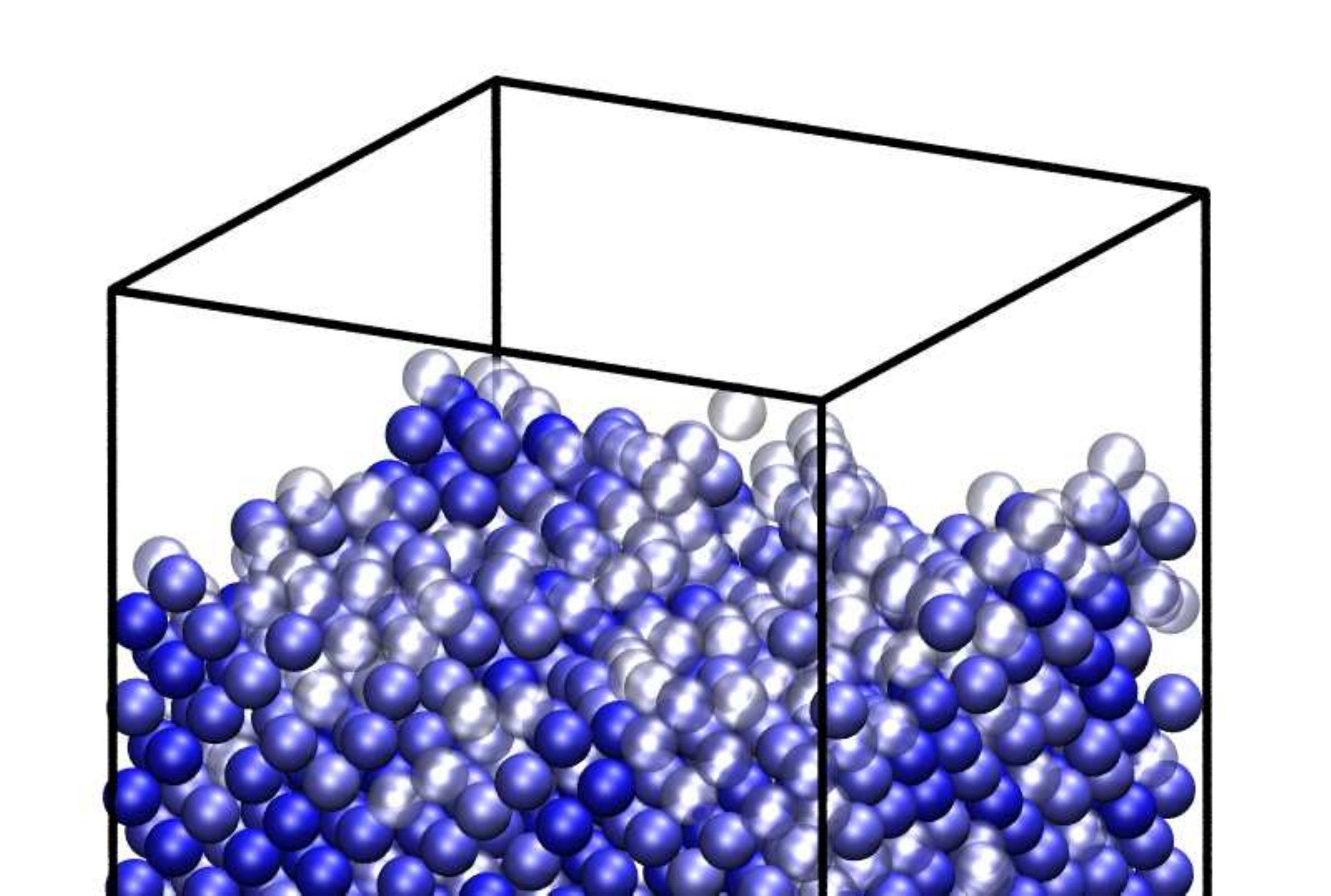}
\caption{
(colour online)
 A snapshot of the solid-liquid interface taken from the 
final part of a well-tempered metadynamics simulation.
The scaled order parameter $\bar{\phi}$ has been used to colour atoms.
The atoms with a liquid-like configuration, with $\bar{\phi}<0.45$ 
have been hidden, the atoms with $\bar{\phi}>0.65$ have been
coloured in blue. Finally, atoms in intermediate 
configurations, with $0.45<\bar{\phi}<0.65$ have been made translucent.
It is clear that - whatever threshold is used to ascertain the 
solid from the liquid state - the interface is not flat on the 
atomic scale.
}
\label{fig:roughness}
\end{figure}

A peculiar feature of our approach is that, at variance with cleavage methods,
the solid-liquid interface is created and ``annihilated'' several times during each 
simulation, so that hysteresis should be much less of a concern. 
When the well-tempered bias is nearly converged, the systems diffuse on
a flattened FES, and the morphology of the interface
corresponds to the most favourable one from a free-energy point of view.
As seen from Figure~\ref{fig:roughness}, such a morphology 
includes a significant amount of roughness at the atomic scale. This might be 
expected from the observation that for the system under consideration
the melting temperature is higher than the roughening transition temperature
for the $(100)$ surface.

\subsection{Analysis of accuracy and system-size effects}\label{sec:error}

Several terms contribute to the error in calculating a complex thermodynamic property
such as \gsl. In actual applications of this method to a real substance one will be concerned 
with the accuracy of the total energy and force model, but this is not an issue in our present proof-of-principle case.
However, there are still two major sources of error we must be concerned with here; namely, a statistical
error stemming from insufficient ergodicity of the sampling (a finite sampling-time error) and 
the inaccuracies caused from insufficient size of the supercell. These finite-size errors introduce a lower bound on the 
acoustic vibrational frequencies, and most importantly might affect the structure of the liquid phase,
introducing spurious correlation that change the liquid entropy thus changing the melting temperature of the
system (although we have seen no evidence for this in the pair correlation function reported above). Moreover, they could in principle induce a strain field in the solid and introduce interactions between the two interfaces.

The finite-sampling error is readily gauged, by performing several independent runs and by checking
how quickly the discrepancy between the reconstructed free-energies converges to zero.
It is shown in Refs.~\cite{welltempered_orig,welltempered_recover} that for simple models the error in the FES, after a
short transient phase when the free-energy basins are being filled,  is expected to decay as the inverse square root of 
simulation time. 

\begin{figure}[bpt]
\includegraphics{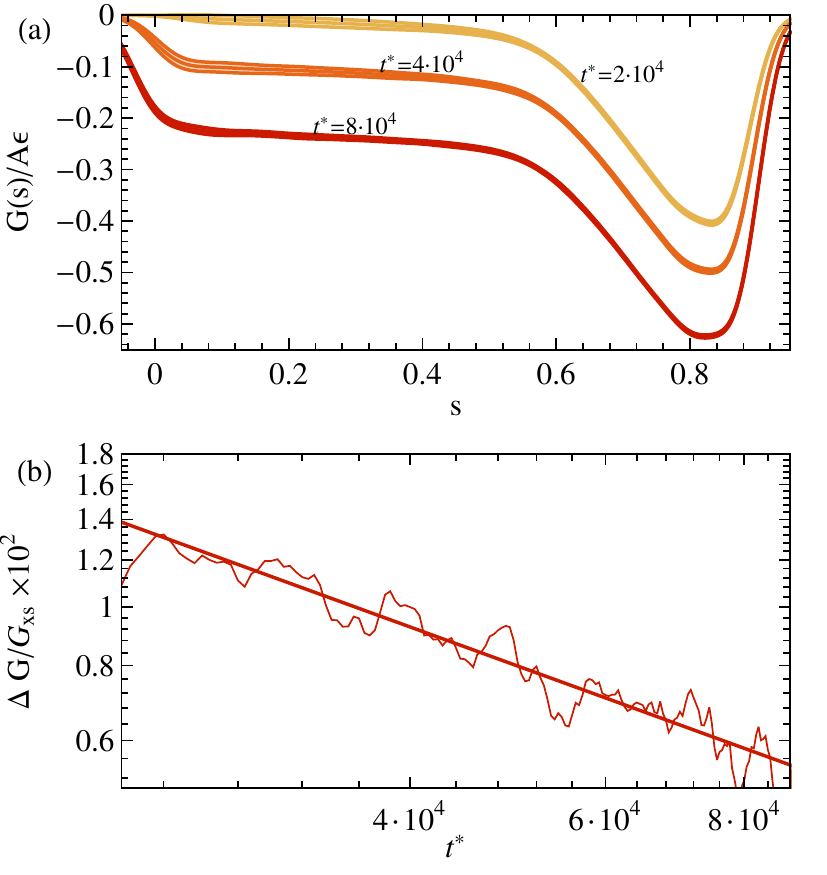}
\caption{(colour online) Convergence of the FES (inset (a)) and its average error (inset (b)) with respect to time for our
1D simulations. Ten simulations have been performed on a $7\times 7\times 12$ supercell, with the 
single-CV setup described in the text. The FESs in inset (a) are constructed by averaging equal-times
biases of the independent runs, and the error-bars correspond to the standard deviation.
In inset (b) we plot such an error, averaged between $\bar{\phi}_l$ and $\bar{\phi}_s$, as a 
function of simulation time. The error is plotted on a log-log scale and the least-square linear fit shows 
that the angular coefficient is close to the theoretical value of $-1/2$ predicted for a simple Langevin model.
}
\label{fig:meta-conv}
\end{figure}

It is reassuring to verify that this behaviour is  also  found in our system, as shown in Figure~\ref{fig:meta-conv}.
This behaviour is to be expected, because for long simulation times well-tempered metadynamics 
corresponds to a histogram-reweighting with a nearly perfect biasing potential. It means also that 
rather than running a very long simulation, one can with equal machine efficiency perform several, shorter, independent
runs, with great advantages from the point of view of parallelisation.

In order to calculate the value of \gsl from the reconstructed FES, we need to monitor in time the estimate 
of the excess free-energy due to the interface,
%
\begin{equation}
G_{xs}(t)=G_{s|l}(t)-G_{s(l)}(t)\rightarrow \gsl A\,,
\label{eq:deltag}
\end{equation}
where we label as $G_{s|l}$ the estimate of the free energy for a configuration with a solid-liquid interface.
As is routinely done in conventional metadynamics simulations, we take as our best estimate
of $G_{xs}$ the incremental average over the final part of the trajectory, well after the 
initial transient:
\begin{equation}
G_{xs}\approx \frac{1}{t_f-t_i} \int_{t_i}^{t_f} G_{xs}(t) \mathrm{d}t\,.
\label{eq:gxc-average}
\end{equation}
We perform ten independent runs, and we can therefore compute an unbiased estimate of the overall statistical error.

\begin{figure}[bpt]
\includegraphics{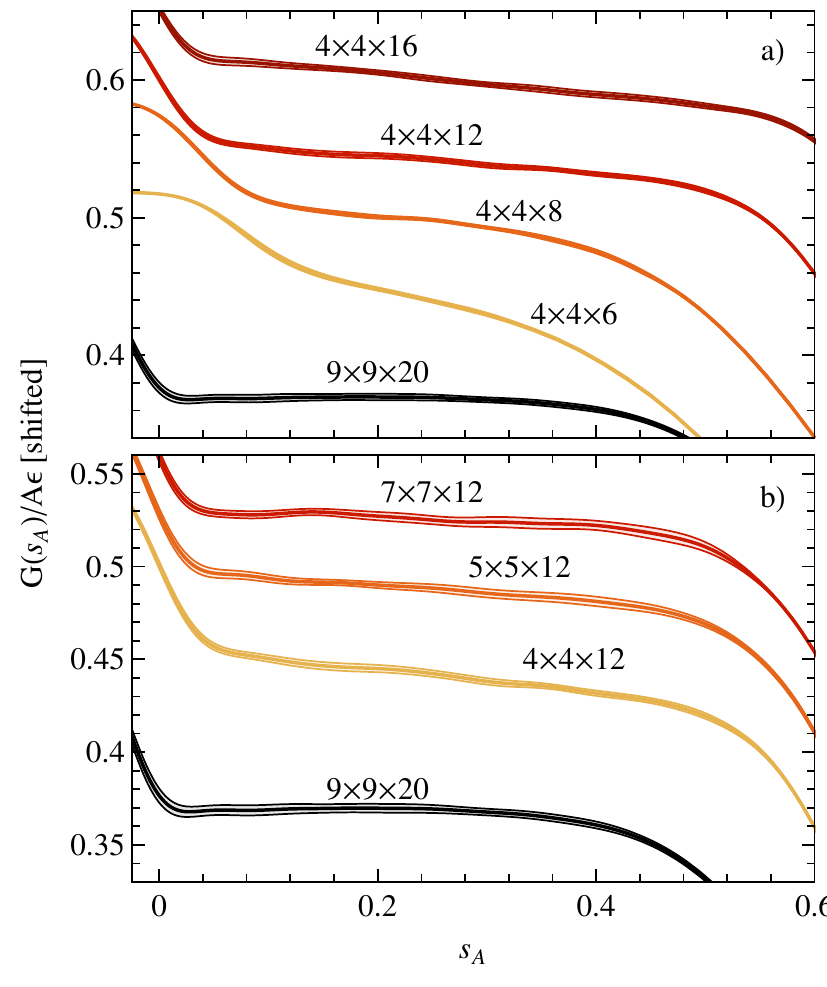}
\caption{
(colour online) (a) The plateau region corresponding to the presence of the solid/liquid interface, with different relative amounts
of the two phases, is drawn for different supercell sizes in the $z$ direction. The curves are shifted
for display purpose, and they are only meant to demonstrate how the flat portion of $G(s)$ becomes
more extended as larger supercells are considered. Finite-size effects are present also in the region 
for $s<0.1$. In inset (b) we compare results for simulations with different in-plane sizes, showing a further
finite-size effect which depends on the change in $T_m$ and causes a residual slope of $G(s_A)/A$ even 
after the complete formation of an interface.
}
\label{fig:meta-size}
\end{figure}

As we previously discussed, in the case of our 2D metadynamics, there are many points on the FES corresponding 
to coexisting solid and liquid phases,  which have the same free energy 
(see Figure~\ref{fig:fes2d}); analogously, an extended plateau region is found in the 1D setup.
Therefore, any point in these regions would be a valid choice for evaluating the interfacial free energy, 
provided that these regions are indeed flat. This brings us to the discussion of finite-size errors.
In fact, at least for a simple, short-ranged potential such as Lennard-Jones, the greatest concern 
is the interaction of the two interfaces along $z$, mediated by the elastic strain field in the solid portion 
and by the altered structure of the liquid in close proximity to the solid/liquid boundary.
Such effects are already clearly evident from the 1D FES reported in Figure~\ref{fig:meta-size}.
In the case of very small supercells ($4\times 4\times 6$ and $4\times 4\times 8$ \emph{fcc} cells, containing 384  and 512 atoms respectively) finite 
size effects are 
quite severe and one can hardly see a plateau region. The free energy at these supercell sizes changes quite rapidly
 over the whole CV space, probably due to the strong interactions between the two interfaces formed during the solid-liquid transition,
which are quite close in such short cells.  As we increase the length of the supercell at constant $xy$ dimensions, from $4\times 4\times 12$ onwards, another feature of the free energy is observed: 
 one can clearly distinguish a plateau with a linear residual slope.  
This can be explained by the reduction in the liquid entropy by the constraint of finite $xy$ dimensions of the supercell, which slightly raises the melting temperature. Since the solid is marginally more stable, once the interface is formed 
there will be a linear increase in free energy as the fraction of liquid phase grows. 
Indeed, simulations for supercells with larger $xy$ dimensions yield a flatter plateau (see Figure~\ref{fig:meta-size}(b)).
The small increase in melting temperature is expected to manifest itself when the width of the supercell
is less than some correlation length $2~L_{corr}$.
$L_{corr}$ can be estimated by looking at the distance at which the pair correlation function (see Figure~\ref{fig:eq-test})
approaches $1$, which in our case is $5\sigma$, corresponding to $\approx 3$ cell parameters, suggesting we need at least  $6\times6$ unit cells in $xy$. Indeed the effect is seen to have vanished by $9\times9$ unit cells (Figure~\ref{fig:meta-size}b).

With these concerns about finite-size effects in mind, we can discuss a reasonable 
protocol to compute $G_{xs}$. For the 2D metadynamics, the region with $0.35<s_A,s_B<0.55$
is sufficiently flat, and we estimate the free-energy of the two-phases configuration
to be  $G_{s|l}=G(s_A=s_B=0.45)$.
Let us now consider the 1D case, where region $B$ is restrained to remain solid.
In this case, due to the finite slope, the estimate of \gsl will depend on which point on the plateau
we take to be corresponding to the value of $G_{s|l}$.    
Hence, we get our best estimate of $G_{s|l}$ as $G(s_A=0.2)$ 
(the point equidistant from the limits of the plateau region)
and estimate roughly the systematic error due to the finite slope as 
$G(s_A=0.3)-G(s_A=0.1)$. 

The results of calculations with different supercell sizes are reported
in Table~\ref{tab:gammasl_vs_size}, where we 
report our best estimate of \gsl and of 
the statistical and finite-size errors($\Delta\gsl^{stat}$ and $\Delta\gsl^{fs}$ respectively).
 $\Delta\gsl^{stat}$ is computed as the root mean square deviation between 
10 independent simulations of $10^7$ timesteps each.
The results of a run using two CVs is also reported for comparison.

\begin{table}

\begin{tabular}{ c c c }
\hline
&
\# atoms (cell)&
\gsl($\Delta\gsl^{stat}$,$\Delta\gsl^{fs}$)\tabularnewline
\hline
\hline

S1 (2D) &
2352 ($7\times 7\times 12$)&
0.37(0.01)\tabularnewline

S2 (1D) &
384 ($4\times 4\times 6$)&
0.39(0.0008,0.02 )\tabularnewline

S3 (1D) &
512 ($4\times 4\times 8$)&
0.390(0.001,0.008)\tabularnewline

S4 (1D) &
768 ($4\times 4\times 12$)&
0.390(0.003,0.005 )\tabularnewline

S5 (1D) &
1024 ($4\times 4\times 16$)&
0.390(0.002,0.006)\tabularnewline

S6 (1D) &
1280 ($4\times 4\times 20$)&
0.390(0.003,0.002)\tabularnewline

S7 (1D) &
1200 ($5\times 5\times 12$)&
0.386(0.003,0.003 )\tabularnewline

S8 (1D) &
2352 ($7\times 7\times 12$)&
0.369(0.002,0.0006)\tabularnewline

S9 (1D) &
6480 ($9\times 9\times 20$)&
0.360(0.003,0.0004)\tabularnewline
\hline
\end{tabular}
\caption{Value of \gsl calculated for different supercell sizes with both 1D and 2D metadynamics. 
The error is reported as $(\Delta\gsl^{stat}$,$\Delta\gsl^{fs}$),  
where $\Delta\gsl^{stat}$ and $\Delta\gsl^{fs}$ are the statistical 
and systematic error respectively, as defined in the text. 
For all sets of parameters, ten independent runs have been performed, each $10^7$ steps long.
\label{tab:gammasl_vs_size}}
\end{table}

\subsection{Comparison with other methods}

With the aid of Table~\ref{tab:gammasl_vs_size} we can now discuss the relative merits of our technique. 
First of all it can be seen that our calculated value for the (100) surface is very close to the ones calculated
by CFM($0.369\pm0.008$) and CM ($(0.371\pm 0.003)$ in Ref.\cite{cleavage_lj} and $(0.34\pm 0.02)$ in Ref.\cite{cleavage_gilmer}).
Although we cannot make such a direct comparison with CNT (because only an averaged value for \gsl, $\gamma_{sl}^{avg}$
 for all possible orientations is given) we point out that their value of $\gamma_{sl}^{avg}=0.302\pm 0.002$ 
is much lower than ours. The anisotropy in \gsl accounts for part of the difference, as the (110) and
(111) surfaces have a lower \gsl \cite{cfm_lj}, but we suggest that the anisotropy is too small to account for all of it.
The fact that the value calculated by CNT is much lower than both ours and that of the CFM and CM may also be due to
the curvature and temperature dependence of \gsl; CNT is the only method dealing with
curved interfaces at temperature below the equilibrium $T_m$,  as noted in
\cite{cnt}. 
We also point out here that we do not neglect the $pV$ term as done in 
the first version of the CM approach\cite{cleavage_gilmer}, but we still recover a free energy higher than that
calculated by CNT. This should rule out the possibility, as supposed in Ref.~\cite{cnt}, that relaxations
 in volume during the formation of the interface could be another explanation for the discrepancies in \gsl. 

In part, the existence of a small discrepancy between results of CM, CFM and the present work, which 
rely on similar thermodynamic assumptions, can be explained in terms of 
differences in the technical details of the calculations.
For instance, in some of the CM calculations temperature-control has been 
implemented by a  non-standard velocity-rescaling method, which might
affect the accuracy of sampling of the canonical ensemble. 
In the present work we have tried to highlight
all the possible sources of statistical and systematic error, to
facilitate further comparison. 
In any case, the discrepancy between different numerical approaches
is negligible when compared to the errors affecting experiments, 
which can give results differing by as much as 300\% (see e.g. Ref \cite{dominique}).
Hence, any of the aforementioned techniques can be extremely
valuable in assisting the interpretation of experimental data and 
the development of new materials.

The small system size required for our simulations will be a particular advantage, since system size is
 by far the biggest limitation in applying more sophisticated potentials.
We obtain reliable results with system sizes as small as about $1000$ atoms, more than two order of 
magnitude smaller than required by both CFMs and CNT. 
CMs require a few thousand  atoms, so the advantage is less impressive. However, we remark that 
the lower bound attainable by CM is most likely set by the need to mitigate hysteresis effects,
while with our metadynamics approach this is not an issue, and the limiting factor here will be the kind
 of interactions between interfaces that are inevitable in all total energy calculations based on periodic boundary conditions.  

In view of the large experimental inaccuracies in the measure of  \gsl (e.g. Ref.~\cite{dominique}) even simple empirical
potentials  would already lead to a quantitative improvement of the knowledge of \gsl. We are applying our method to some of these cases 
and the results will be published in a future paper. Nevertheless, it may be that  the accuracy or information given by a self-consistent electronic structure method is desired for the interfacial free energy calculation, in which case our approach would still be a promising candidate.  
With high performance computers, simulating a few hundred atoms for several hundred picoseconds is within the reach of present,
widely used molecular dynamics methods employing so-called
\emph{ab initio} (electronic density functional) techniques for the calculation of interatomic forces.
This would probably result in 
better predictive power and smaller overall errors despite the possibility of mild finite-size effects.

With a view to performing calculations with more sophisticated potentials, metadynamics
offers a further advantage over the other techniques. One could implement
a process of iterative refining, whereby one performs
a sequential set of calculations with potentials of increasing sophistication and computational cost, 
in order to reduce the burden of levelling the FES.
 In fact, the major features of the FES can be captured by the use of very simple potentials reproducing the 
nearest neighbour bonding in the real material. 
This first level FES, $G^{(0)}(s)$,  could then be used as the initial bias for a second
metadynamics run, to be performed with a more accurate (and expensive) potential. At this stage,
one will have the much easier task of correcting the discrepancy between $G^{(0)}(s)$
and the FES of the new potential, $G^{(1)}(s)$. 
This scheme could be repeated with increasingly accurate potentials.

\section{Conclusions}

In the present paper we have presented a novel approach to the calculation of the 
solid-liquid interface free energy \gsl, and discussed its application to
the calculation of \gsl for the $(100)$ surface of a Lennard-Jones solid in
contact with its liquid.
Our method is based on the definition of a new order parameter, which is designed to 
identify \emph{fcc}-ordering of atoms in the orientation of choice, compatible with the periodic 
boundary conditions, and uses metadynamics simulations to estimate free-energy differences
between the bulk phases and configurations where macroscopically flat interfaces are 
present.
We obtain results for \gsl in good agreement with previously proposed methods. 
Moreover, our technique offers several advantages compared to previous approaches discussed
in the literature. It requires fewer atoms than those
methods based on macroscale approximations, such as measuring capillary fluctuations
or the critical nucleation radius, while being less affected by hysteresis than 
cleavage methods, since the interface is created and destroyed several times during each 
simulation as equiprobable sampling of the free energy surface is approached. 
We discuss at length the different sources of error, and how they can be controlled.
In particular, we show that our approach is effective even for supercells containing fewer than 1000 atoms, with finite-size errors whose importance can 
be gauged easily. For this reason, we speculate that it would be possible to
perform an \textit{ab initio} calculation of \gsl, at the level of electronic density functional theory. To this end, we suggest that an iterative refinement
scheme, which starts with a biased free-energy surface computed from a semi-empirical
potential, could be a helpful starting point for obtaining converged results within reasonable
computational time. We plan to attempt these calculations in the near future,
after having further validated our method by comparison with experiments in the case of a simple metal
for which we expect empirical potentials to be adequate.


\section{Acknowledgments}

The authors thank Alessandro Laio for very fruitful discussions
about metadynamics and Mark Asta for an early reading of this manuscript
and his valuable comments.
We would also like to thank Alessandro Barducci, Max Bonomi and 
Paolo Raiteri for help with PLUMED and advice 
about the subtleties of metadynamics. Finally, we gratefully acknowledge 
the COST Action P19 (Multiscale Modeling of Materials) for travel
funding that allowed the collaboration between the authors, and
 EPSRC for support under Grant No. EP/D04619X.

\bibliographystyle{achemso}
\providecommand{\refin}[1]{\\ \textbf{Referenced in:} #1}

\end{document}